\documentclass[11pt,twoside]{article}

\usepackage{asp2006}
\usepackage{epsf}
\usepackage{lscape}
\usepackage{graphicx}

\begin{document}
\title{Outflows from Massive Stars}   
\author{John Bally}   
\affil{Department of Astrophysical and Planetary Sciences, \\ 
University of Colorado, Boulder,   CO 80389, USA.}    

\begin{abstract}
The properties of outflows powered by massive stars are reviewed 
with an emphasis on the nearest examples, Orion and Cepheus-A.  
\end{abstract}

\section{Introduction}   

Outflows are  ubiquitous in star forming regions.  They  provide observable
constraints on the accretion histories and dynamical  evolution of forming 
stars, and represent a major channel for self-regulation of star
formation and feedback.
Our understanding of outflow phenomena has grown dramatically 
as new wavelength regimes, higher resolution, and greater sensitivity
produced a flood of new observations.    However,  observers have
mostly focused on relatively nearby  outflows that are powered by 
low- to moderate-mass YSOs (for reviews  see Reipurth \& Bally 2001; 
and  the articles in  Reipurth, Jewitt, \& Keil 2007).   Thus, the study of outflows
from massive stars is still in its infancy.  

Massive stars 
are rare and tend to form in high pressure, opaque  cores that produce
clusters.   Therefore, massive star outflows tend to be distant, highly 
obscured,  and located in crowded and confused environments.  Furthermore, 
their swept-up molecular shells and cavities can be  rapidly obliterated by UV radiation 
and winds as the massive stars reach the main  sequence.    The nearest 
massive star outflow emerges from the highly obscured  BNKL complex in the 
OMC1 cloud core  located immediately behind the  Orion Nebula at a distance 
of about 400 to 430 pc (Menten et al 2007; Hirota et al 2007; Sandstrom et al. 2007) .  
The next best studied massive-star outflow is associated with  Cepheus-A  
located about 725 pc from the Sun.  Following  a few general comments about 
massive star outflows,  the OMC1 and Cep-A flows will be discussed in some detail.

 \section{Outflows from Massive Young Stellar Objects (MYSOs)}   
 
MYSOs can be identified by their large IR luminosities,  the presence of maser 
emission, ultra- or hyper-compact HII  regions,  massive hot cores,  high velocity 
wings in mm or sub-mm emission lines,  the presence of shock-excited species
such as H$_2$ and [FeII],  and, if the obscuration is sufficiently low, 
Herbig-Haro (HH) objects. 
MYSOs can be subdivided according to their luminosity.  Intermediate mass objects
that reach peak luminosities of   L $< 10^3$  L$ _{\odot}$  become Herbig Ae or Be 
stars with masses of 2 to 8 M$_{\odot}$ and are  massive analogs to T Tauri stars.   
They produce copious soft-UV, but little or no ionization of hydrogen.  MYSOs with  
peak luminosities  L = $10^3$ to  about $3  \times 10^4$  L$ _{\odot}$ become  early 
B or late  O  stars that can rapidly  photo-ionize their environments.    Most MYSOs
(including Cep A)  are in these two  categories;  they tend to have outflows that are 
scaled-up versions  of  outflows from low-mass stars.  Their outflows tend to be 
highly collimated,  contain multiple internal working surfaces, have  mass-loss rates 
of $10^{-5}$ to $few \times 10^{-4}$ M$_{\odot}$~yr$^{-1}$, 
momentum injection rates  around 
$10^{-4}$ to $10^{-2}$ M$_{\odot}$~yr$^{-1}$~km~s$^{-1}$, 
and mechanical luminosities  of $10^{-1}$ to $10^2$~L$_{\odot}$ that increase 
approximately  linearly with stellar luminosity.  MYSOs  that approach or exceed
L $\approx 10^5$ ~L$_{\odot}$ become the exceedingly rare, most massive early 
O-type stars.   Their outflows are frequently poorly collimated with a morphology 
resembling an explosion; examples include the OMC1 / BNKL outflow in Orion 
and G34.26+0.15. 

Beuther and Shepherd (2005)  proposed an evolutionary scheme for massive YSOs. 
As the most massive stars accumulate their mass, they increase their luminosity and
pass through each of the luminosity classes above.   MYSOs with luminosities
indicating spectral type B or later tend to have highly collimated outflows.  
By the time such stars have  acquired a mass of about 10 M$_{\odot}$, they will 
have reached the main sequence and their UV radiation fields will heat and
ionize their immediate surroundings.  MYSOs with luminosities of early B or late
O stars tend to be surrounded by hyper- or ultra-compact HII regions and hot 
molecular cores with complex chemistry and their outflows tend to be less-well 
collimated.  The most luminous MYSOs sometimes
have very poorly collimated flows.  However, except for Orion, these luminous 
objects tend to be studied with poor spatial resolution.  Thus, their morphology
may not be fully resolved in available data; some may be powered by many individual YSOs.  
Some  ultra-massive outflows such as DR21 are likely to be powered by the collective 
effects of entire clusters of stars.   

In luminous sources,  UV radiation and line-driven winds can make major 
contributions to the momentum and energy injected into surrounding gas.    
When large reservoirs of dense, molecular gas
survive in the immediate vicinity of a massive star or cluster, Lyman continuum can produce
a compact HII region that vents into the lower density ISM.   For certain geometries such
as thick torii, or for a massive star embedded in a cylindrical cavity within a dense cloud, 
the ionization fronts will be confined by recombinations to a value comparable to
the radius of the cavity.  Thermal expansion of the high pressure, photo-ionized plasma 
ablating from the walls of the cavity close to the star, will drive a flow along the axis of the cavity.   
The plasma can sometimes flow through a recombination front at about  
the sound speed in ionized gas, recombine, and cool to
form a hypersonic flow of HI.  A quasi-static flow can be established as photo-ablation
from the walls of the reservoir replenishes the lost plasma. Thus, the HII region remains 
confined and does not evolve on a dynamical time, but can effectively fuel a powerful,
radiation driven outflow.   Such an ionizing radiation driven model may explain the energetics 
of the most massive  outflows  such as the 200 M$_{\odot}$ Mon R2 outflows and the 
3,000 M$_{\odot}$  DR 21 HI outflow (Russell et al. 1992).

Massive stars tend to form in clustered environments where many 
YSOs power outflows simultaneously and where the gas pressure and turbulent
energy density can be very large compared to lower-mass star forming regions.    
Figure 1 illustrates mm-wave molecular emission from the outflow complex  
emerging from the cluster associated  with IRAS 05358+3543 located about 
2 kpc from the Sun in the Auriga star forming clouds (Beuther et al. 2002).  At
least five distinct flows can be seen.    The  2.12 $\mu$m near-IR  emission 
line of H$_2$ traces shock-excited molecular gas.   Several additional outflows
are revealed by this tracer (Figure 2).  Some outflows from massive stars show 
evidence for large changes in the orientation
of the outflow axis on a time-scale comparable to the outflow dynamical age.  Examples 
include IRAS 20126+4104, and Cep-A.   While C-symmetric bends in outflows are
generally an indication of relative motion between the star and surrounding medium
(e.g. Bally et al. 2006), point symmetry (S- or Z-shaped outflows) are generally an indication
of a precessing outflow. 

\begin{figure}[!ht]
\includegraphics[width=0.9\textwidth]{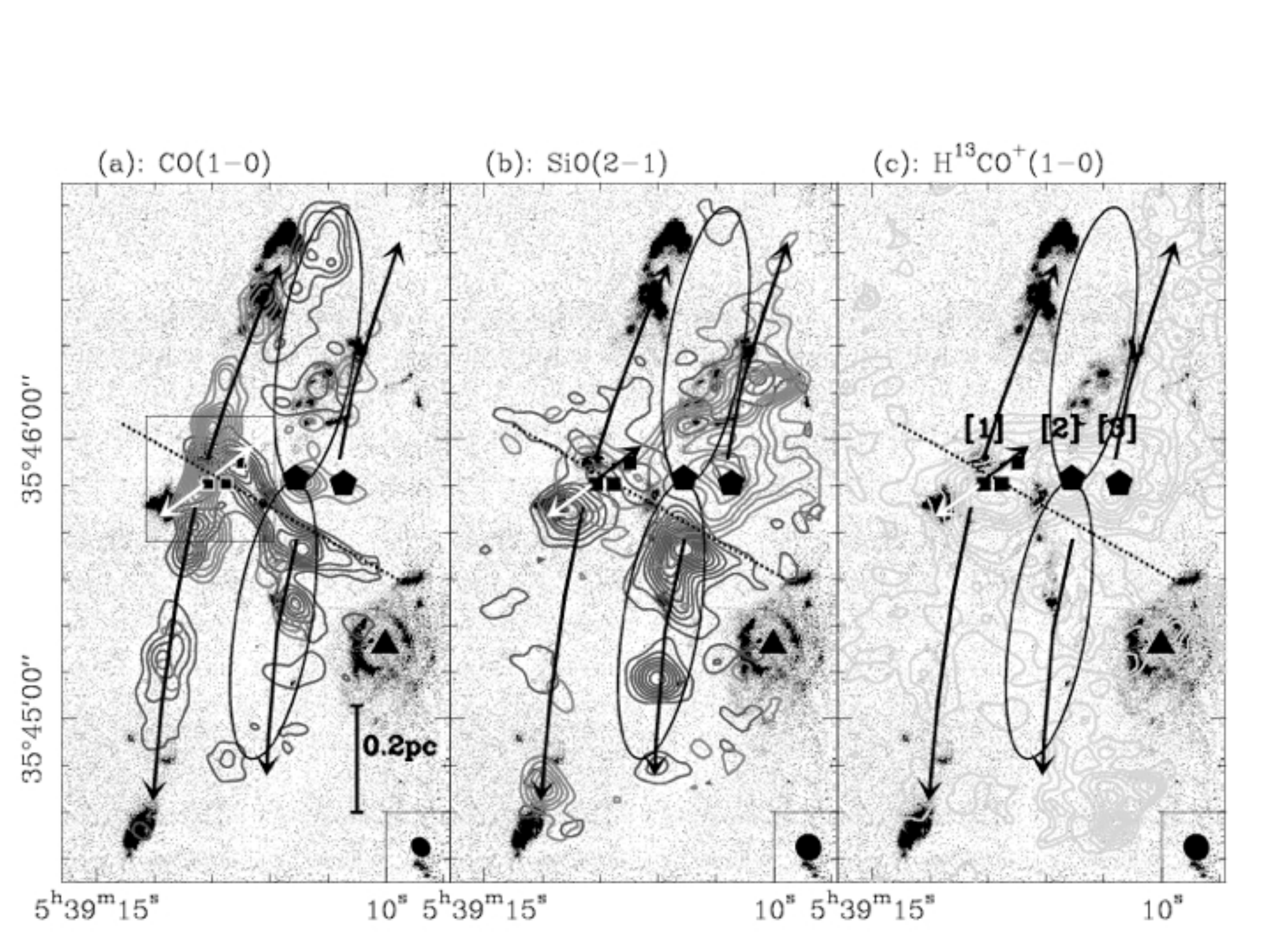}
\caption{Multiple outflows bursting from the cluster associated with IRAS 05358+3543 
(Beuther et al. 2002).
}
\end{figure}

\begin{figure}[!ht]
\includegraphics[width=0.8\textwidth]{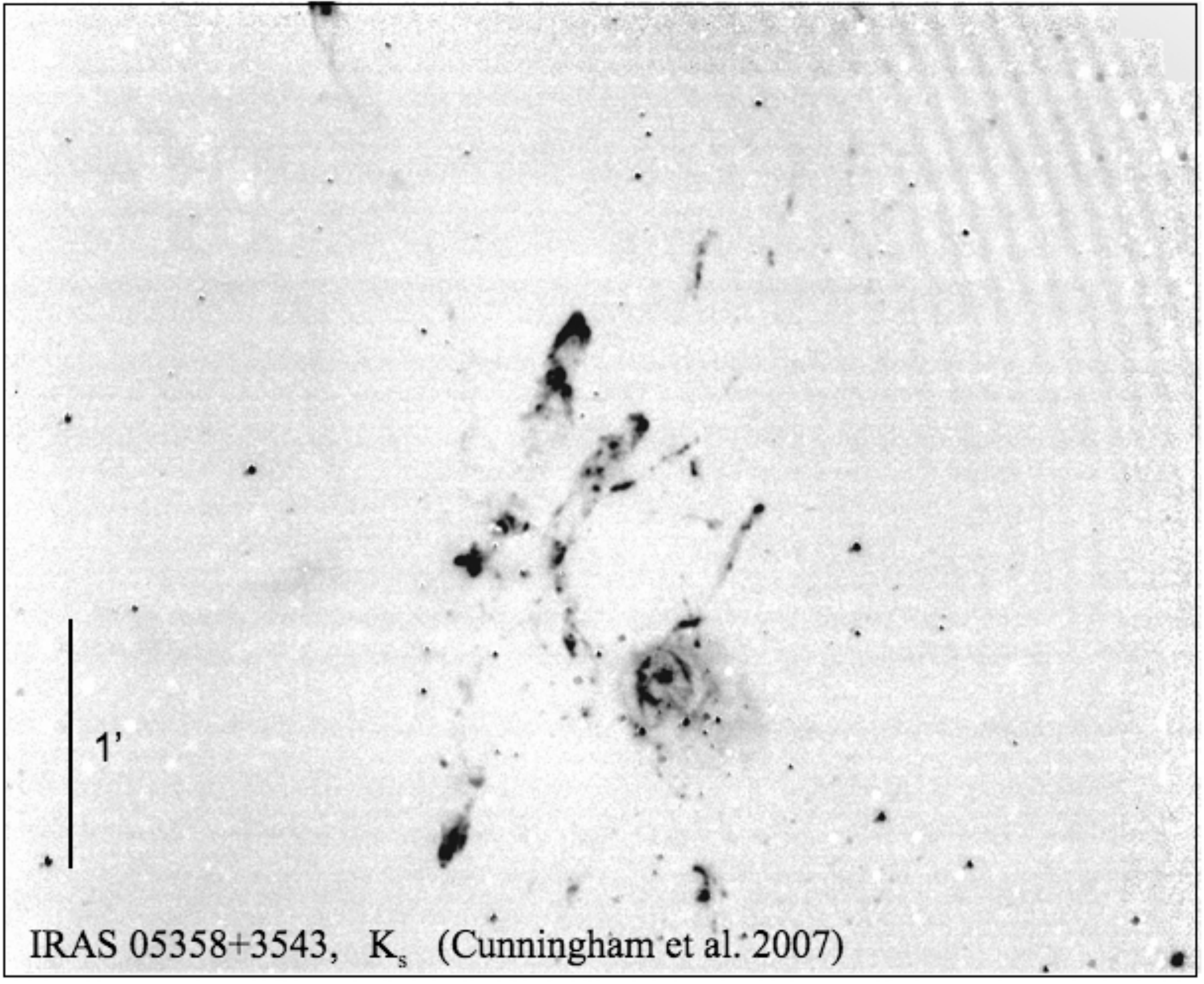}
\caption{Multiple jets and additional outflows emerging from IRAS 05358+3543 and
the associated young cluster as revealed by a 2.12 $\mu$m, continuum subtracted
narrow-band image obtained with the near-IR camera NICFPS on the Apache Point 
Observatory 3.5 m telescope using  0.5\% pass-band filters. 
}
\end{figure}

\section{The Cepheus A Outflow}

The Cep A star forming complex contains the second nearest region of massive star formation,
next to the Orion complex.   Located at a  distance of 725 pc  (Blaauw et al. 1959; Crawford \& 
Barnes 1970),   the Cep OB3 association contains a 20 by 60 pc molecular cloud that contains 
six localized peaks of CO emission designated Cepheus A through F (Sargent 1977, 1979).  
Cep A  contains dense molecular clumps (Torrelles et al. 1993), molecular outflows 
(Rodriguez et al. 1980,  Narayanan \& Walker 1996), H$_2$O and OH masers  (Cohen et al. 1984), hyper-compact H II regions  (Hughes  \& Wouterloot 1984), variable radio continuum 
sources (Garay et al. 1996), Herbig-Haro objects  (Hartigan  et al. 1986), 
bright shock-excited H$_2$ emission (Hartigan et al.  2000), a cluster of far-infrared sources
with a luminosity of  $2.5 \times 10^4$ L$_{\odot}$  (Koppenaal et al. 1979), and a cluster of 
Class I  and Class II  YSOs (Gutermuth et al. 2005).  The bulk of the luminosity arises from 
radio sources HW2 and HW3c (Hughes \& Wouterloot 1984) that are associated with bright 
H$_2$O masers. 

Rodriguez et al. (1980) found a massive bipolar molecular outflow aligned primarily  east-west.
Single dish CO maps with 15$\arcsec$ resolution (Bally \& Lane 1990; Torrelles et al. 1993;
Narayanan  \& Walker 1996) show  that the flow is complex with a second redshifted peak 
west of the core, and with additional components  aligned northeast--southwest.    
The inner part of the flow is blueshifted  east and redshifted west of the radio source HW2.   
The  additional  redshifted component 
appears 6$\arcmin$  west of HW2, bends to the northwest, and  can be traced up to 
12$\arcmin$ from the Cep A core.   The central 2$\arcmin$ region contains high velocity (HV) 
as well as more compact extremely high velocity (EHV) CO components with a radial 
velocities ranging from -50 to  70  km~s$^{-1}$ relative to the CO centroid. 
The axis  of the EHV outflow  is rotated roughly 40$^{\circ}$ clockwise relative 
to the HV outflow on the plane of the sky. The smaller extent together with the higher velocity 
suggests that the EHV flow traces  a younger  outflow component.  Several self-absorption dips 
follow trends seen in the low-velocity  line wings, with  regions east of HW2 blueshifted and 
west of HW2 redshifted.  Thus  cooler,  self-absorbing  gas traces the low-velocity bipolar outflow, 
rather than  stationary gas surrounding the outflow.   At low velocities, there are additional 
blue- and redshifted components centered on HW2 that are oriented northeast--southwest.

The Cep-A outflow complex contains several Herbig-Haro objects, including the extremely bright HH~168 located about 90$\arcsec$ due west of HW2, and several fainter bow shocks 
located to the east (Hartigan, Morse, \& Bally 2000).     Fainter HH objects (HH~169 and 174) are
located in the eastern, blueshifted lobe.   Near-IR images show an extremely bright 
reflection nebula centered on HW2 with an illumination cone that opens towards the northeast.
The 2.12 $\mu$m H$_2$ line  (Figure 3) exhibits a complex, filamentary structure.  

\subsection{Bipolar outflow from HW3c: HH~168 and its Counterflow}

A series of bright arcs and bow shocks located 1 to 3$\arcmin$ west of HW2 are 
associated with HH~168. The axes of symmetry of most of these shocks indicate 
a point of origin about 20$\arcsec$ to 30$\arcsec$ south of HW2.   The luminous IR, maser, 
and radio source HW3c lies near this axis.  A dim, 40$\arcsec$ diameter bow shock
is located on this axis about 2$\arcmin$ east of HW3c, directly opposite HH~168.   
This feature and the visual-wavelength  part of HH~168 are symmetrically placed about HW3c.
Thus, HW3c is likely to be the source of these shocks and the associated CO 
outflow components. High-resolution cm-wavelength VLA observations (Garay et al 1996) 
reveal a chain of radio sources approximately aligned with the axis of this flow, 
and may either trace an ionized jet, or the interface between a jet and surrounding 
dense cloud material.   Faint arcs of emission about 1$\arcmin$ north of HH~168 
trace the edges of a globule that contains an embedded, low-luminosity  star.  This 
emission is likely to be excited by UV radiation from the Cep OB3 association and is 
thus probably fluorescent in nature.

\begin{figure}[!ht]
\includegraphics[width=1.0\textwidth]{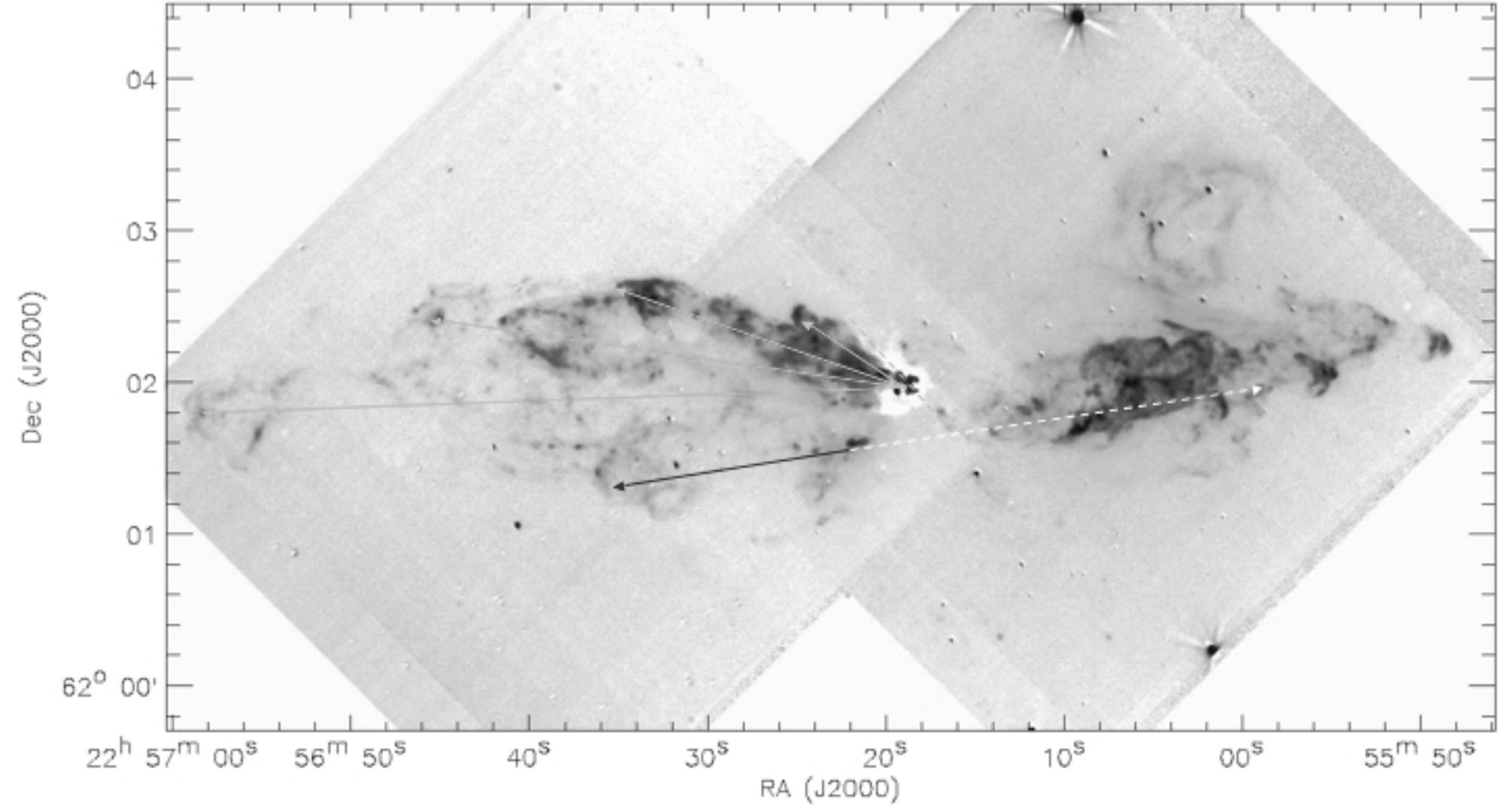}
\caption{Continuum subtracted near-IR 2.12 $\mu$m emission form 
shock-excited H$_2$ in the Cep-A outflow complex.   As discussed in the 
text, HW2 (located at the base of the 4 grey arrows inside the white patch near
the middle of the image) appears to drive a pulsed, precessing
jet whose orientation changes may be driven by forced precession induced by a companion
star in an eccentric, non-coplanar orbit.  The various suspected ejection directions
are indicated by the light gray lines.  The oldest (east-facing) ejection appears to power
HH~174 located at the eastern edge of the image.   The next two ejections may be
responsible for HH~169 located near the northern boundary of the eastern outflow lobe. 
The current orientation of the radio jet emerging from HW2 is at position angle (PA) = 45$^o$
(northeast--southwest; dashed grey line).
The bright objects HH~168 (dashed white line) west of the core, and the H$_2$ bow 
(near the tip of the solid black line) may be powered by HW3c located due south of
HW2 along the dashed white line. From Cunningham (2006).}
\end{figure}

\subsection{A pulsed, precessing Outflow from HW2?}

The blueshifted, eastern lobe of the Cep-A outflow contains four distinct chains of 
H$_2$ emission radiating away from the immediate vicinity of HW2, each of which 
terminates in well-formed bow shocks.  The axes of these 
chains, defined by lines connecting HW2 to the bow shocks at the eastern and north-eastern
ends of the chains,  shift systematically clockwise from nearly  east--west  to northeast--southwest.  
The longest chain, which terminates at HH~174, has a position angle measured 
form north to east of PA $\sim$ 95$^o$.   The second chain, which terminates at the
eastern component of HH~169, has PA $\sim$ 80$^o$.  The third chain terminates
at the western component of HH~169, has  PA $\sim$ 65$^o$.  The fourth chain ends in a bright
but compact H$_2$ bow at PA $\sim$ 55$^o$. The current orientation of the HW2 radio jet
is  PA $\sim$ 45$^o$.     The chains of H$_2$ knots get progressively shorter with 
decreasing PA. This remarkable progression may be an indication that HW2
powers a pulsed and precessing jet.

A rough dynamical age for each chain can be estimated by  dividing the length of each 
chain by the velocity of its tip.   The radial velocities of the HH objects located at the
ends of the first three chains are low ($V < 60$ km~s$^{-1}$, indicating that 
they most likely are moving close to the plane of the sky.    Fabry-Perot imaging of the H$_2$ 
emission (Hirart et al. 2004) also indicate low, but chaotic radial velocities.  
On the other hand, excitation of the visual wavelength H$\alpha$ and [SII] emission of the eastern
HH objects   and comparison with the speeds of other HH objects located at similar
distances from their sources, indicate that shock speeds of order 100 km~s$^{-1}$ are
not unreasonable.      Using this speed, the dynamical ages of the four eastern H$_2$
chains are 10200, 7300, 4400, and 2100 years, respectively, indicating that an eruption /
ejection event  occurs approximately every 2,200.  Furthermore, the presence of the HW2
radio jet indicates that there is currently an eruption underway.

Moeckel \& Bally (2006, 2007a, b) use SPH simulations to show that in a dense proto-cluster,
massive stars surrounded by circumstellar disks have a relatively high probability of capturing
a passing cluster member because the disk makes such encounters dissipative.    While such
capture processes are relatively unimportant for Solar mass stars,  they can be very important
for massive ones due to their relatively much larger gravitational radii.    Generally, 
binaries formed by such a capture process will have secondaries that move in eccentric and 
highly inclined orbits with respect to the plane of the disk or rotation
axis of the primary.    The motion of a companion on an eccentric and inclined orbit can drive
forced precession of the disk.  In contrast,  multiples formed from disk or core fragmentation 
tend to be move in co-planar orbits which do not result in precesing disks.  Furthermore,
the periastron passage of the companion can result in major disk perturbations that can
result in episodes of mass accretion from the inner disk onto the primary.

Moeckel \& Bally (2007a) modeled encounters between a 20 M$_{\odot}$
star with a 2 M$_{\odot}$, 500 AU disk and impactors of varying masses, periastron, and 
inclination angles. Cunningham, Moeckel, \& Bally (2008) scaled those simulations to a 
mass similar to HW2, a 15 M$_{\odot}$  primary with a 350 AU, 1.5 M$_{\odot}$ 
disk.  Disks were modeled using  1.28 $times$ 10$^5$ particles, and followed through 5
encounters with the captured secondary.  The mass of the impactor was chosen to 
be 5 M$_{\odot}$ because a fairly massive 
companion is needed to torque the disk through $\sim 10^o$ during a periastron passage. 
The SPH simulations show that massive partners are more likely to be captured than
low mass ones.   Disks were modeled using  $\sim 1.28 \times 10^{5}$ particles, and 
followed through 5 encounters.  These models show that a capture formed binary with 
a secondary having about 1/3 of  the mass of  the primary, orbiting in a 2,000 year 
orbit can cause a massive disk to precess by about the amount observed in Cep A.

\section{The Orion OMC1 Outflow}

The BN/KL complex contains a remarkable, wide opening-angle ($>$ 2 radian) 
outflow traced by molecules such as CO and  NH$_3$ that exhibit broad 
($>$ 100 km s$^{-1}$) emission lines (Kwan \& Scoville 1976; Wiseman \& Ho 1996)
and high-velocity OH, H$_2$O, and SiO maser emission (Genzel et al. 1981;
Greenhill et al. 1998).  Hundreds of individual bow shocks are seen 
in the near-infrared wavelength emission lines of [FeII] and H$_2$ 
(Kaifu 2000; Figure 4).   A few of these shocks protrude from the 
molecular cloud and can be seen at visual wavelengths.  The morphology 
suggests an explosion; proper motions of visual and near-infrared features 
indicate a common age of about $10^3$ years or less for many features
(Doi et al. 2002; Figure 5).  The outflow contains about 8 M$_{\odot}$ of 
accelerated gas with a median velocity of about 19 km s$^{-1}$ and maximum 
speeds in excess of  400  km s$^{-1}$.  By one estimate, the current 
lower bound on the momentum and kinetic energy content are about 
$160$ M$_{\odot}$ km s$^{-1}$ and $4 \times 10^{46}$ ergs, respectively
(Snell et al. 1983).  Other estimates indicate an outflow kinetic energy 
as high as $4 \times 10^{47}$ ergs (Kaifu et al. 2000).   The OMC1 core 
has a bolometric luminosity estimated to be about $10^5$ L$_{\odot}$ 
(Gezari, Beckman, \& Werner 1998), indicating that the outflow region must 
contain one or more massive stars.

Multi-epoch radio-frequency images show that the three brightest 
radio-emitting stars in OMC1, sources BN, I, and n, have proper motions 
(motions in the plane of the sky) of 26, 15, and 24 km s$^{-1}$ away from 
a region less than 500 AU in diameter from which they were ejected about 
500 years ago (Rodriguez et al. 2005; Gomez et al. 2005).  Apparently, 
a non-hierarchical multiple star system containing at least 4 massive 
members experienced a dynamical interaction resulting either in the 
formation of a tight binary or possibly a stellar merger (Bally \& Zinnecker
2005) whose (negative) gravitational binding energy ejected these stars 
from the OMC1 core.  With estimated stellar masses of 10, 20, and 10 
M$_{\odot}$ for BN, I, and n respectively, the kinetic energy of the 
stars is $2 \times 10^{47}$ ergs (Rodriguez, et al., 2005; Gomez et al. 2005).   
Thus, the total energy required to eject the stars and drive the OMC1 
outflow is about $10^{48}$ ergs.  This energy must be generated by the 
infall of two or more stars into a deeper gravitational potential well.   
Assuming that source I is a binary containing two 10 M$_{\odot}$ stars, 
its members must be separated by less than 11 AU, the orbital period must 
be shorter than 7 years, and the perihelion velocity of the stars must 
be at least 70 km s$^{-1}$.  This stellar ejection expelled a total 
of about 50 M$_{\odot}$ of stars and gas from the OMC1 core.

The point of origin of the ejected stars is located at the center of 
the BN/KL outflow about 4$\arcsec$ (2000 AU) northwest of the present location 
of source I. The stars and outflow have similar ages and kinetic energies.  
Therefore, it is likely that the outflow and dynamical ejection of the 
stars have a related origin.  During  the formation of a tightly bound, 
massive binary in a small-N interaction, gravitational potential energy can 
be transferred to the other stars, but their motion has little effect 
on the surrounding gas.  

\begin{figure}[!ht]
\includegraphics[width=0.8\textwidth]{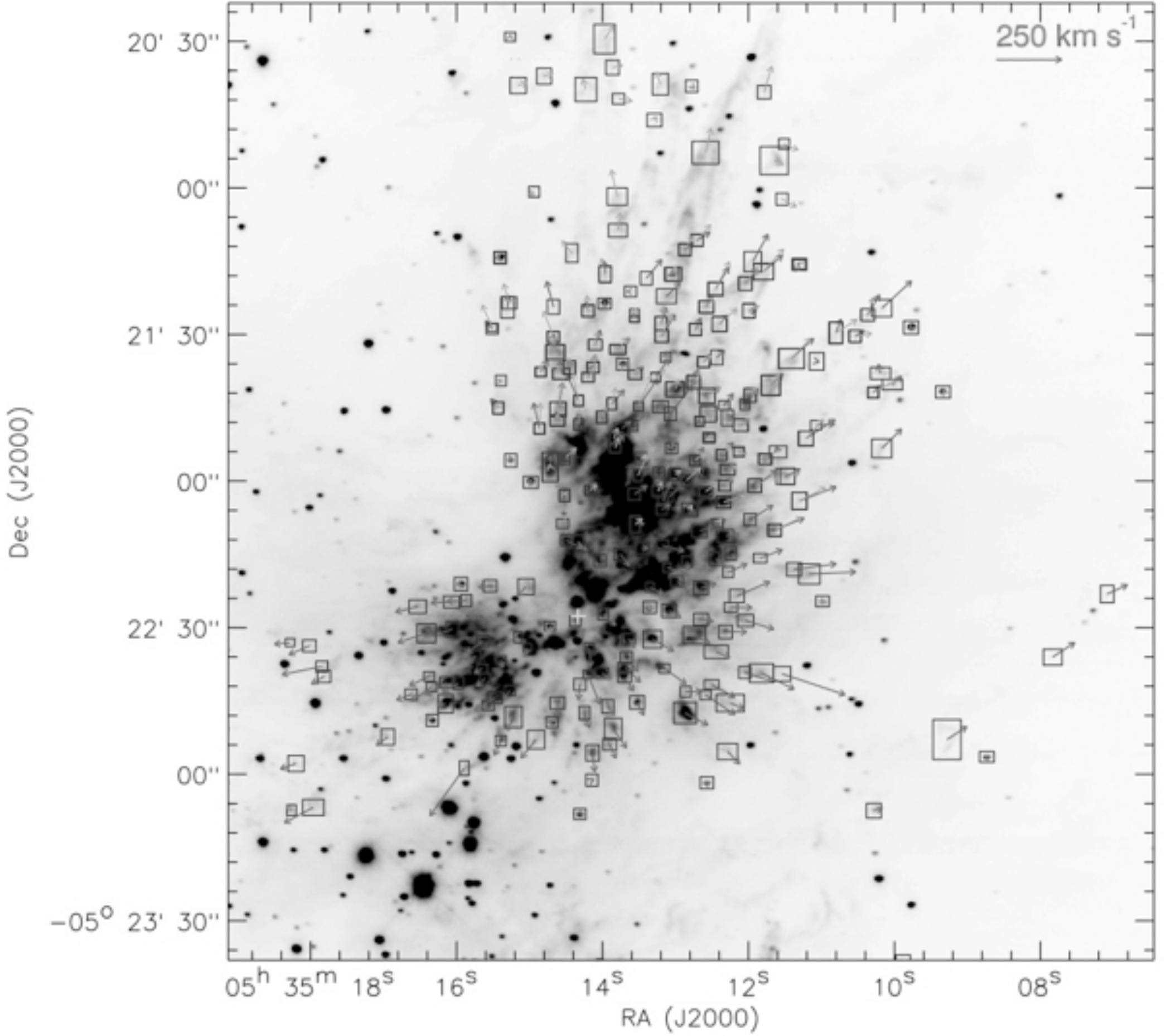}
\caption{The OMC1 outflow as observed in the 2.12 $\mu$m lines of
H$_2$ using a 0.5\% passband filter and the NICFPS camera on the
Apache Point Observatory 3.5 meter telescope.  Vectors show
showing proper motion measured by comparing this image, taken in 2005,
with the Allen \& Burton (1993) discovery images and the Kaifu et al.
(2000) image obtained with the Subaru 8.4 meter.  Taken from 
Cunningham (2006).
}
\end{figure}

\begin{figure}[!ht]
\includegraphics[width=1.0\textwidth]{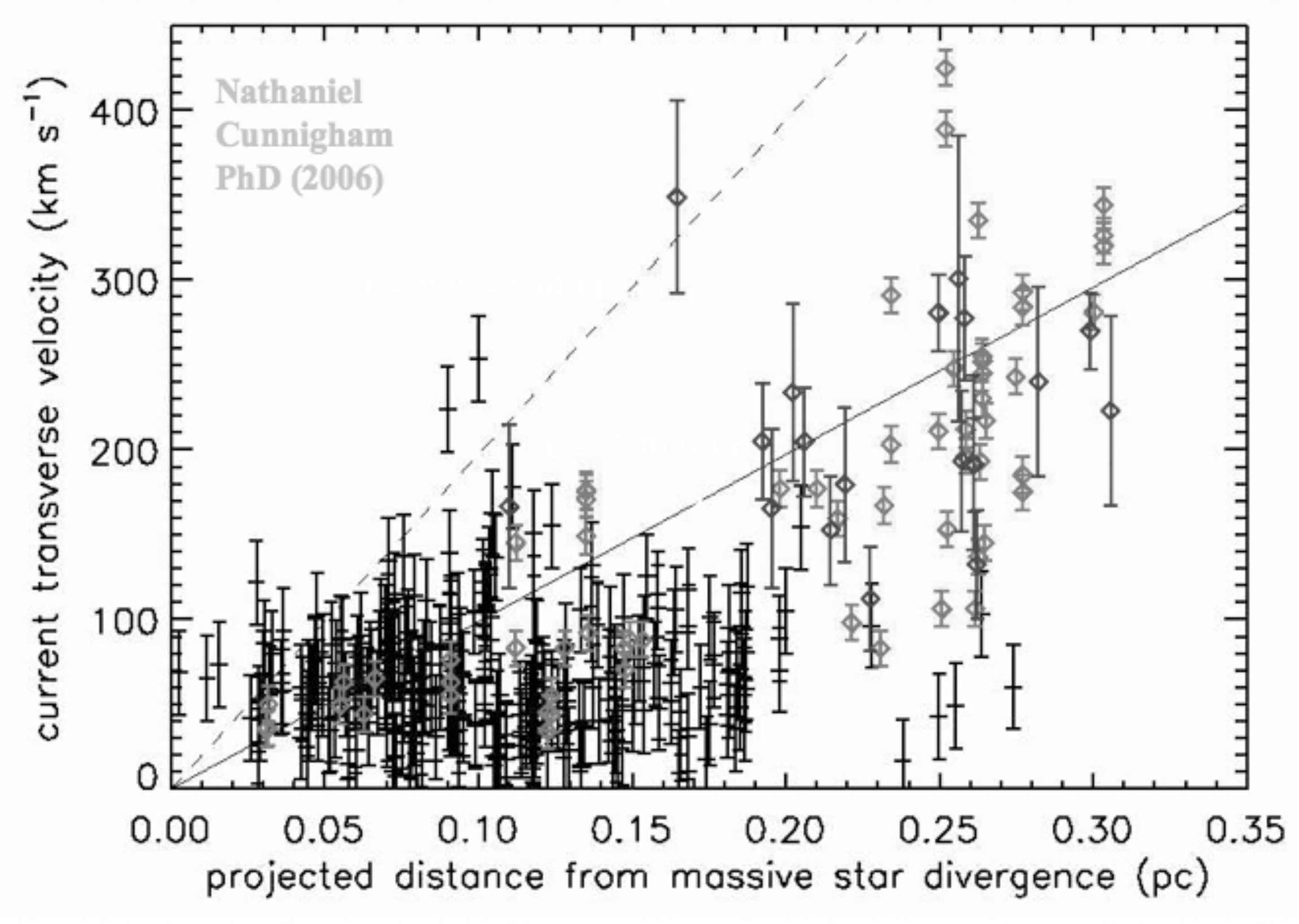}
\caption{Proper motions in the OMC1 outflow as observed in the 2.12 $\mu$m lines of
H$_2$ (black) and at visual wavelength using narrow-band filters on the
Hubble Space Telescope (grey).  From Doi et al. ( 2002) and Cunningham (2006).
}
\end{figure}

\subsection{The Magnetic Bomb}

We propose that over the last $10^5$ years, the birth and subsequent 
orbital motion of a non-hierarchical cluster of 4 or 5 massive stars 
in OMC1 led to the amplification of the ambient magnetic field until 
it stored about $10^{48}$ ergs of magnetic energy in the volume inhabited 
by the cluster.  About 500 years ago, this cluster decayed into a 
hierarchical configuration consisting of a short-period run-away 
binary and two high-velocity stars.  The removal of about 50 M$_{\odot}$ 
left the circum-cluster gas in the region severely magnetically 
over-pressured.  The sudden release of magnetic energy, in effect a 
`magnetic bomb' (Matt, Frank, \& Blackman 2004, 2006; 
Lynden-Bell 1996;  Wheeler, Meier, \& Wilson 2002), may have been 
responsible for launching the OMC1 outflow.  The following paragraphs 
explain in more detail how this scenario may work.

As a cloud core collapses, interstellar magnetic fields will be 
dragged in with the gas that eventually forms a small group of 
massive stars.  The resulting magnetic field strength grows as 
the field lines are compressed, wound, and stretched and is thought 
to scale as $n_H^{\alpha}$,  where $n_H$ is the number density of 
the gas and $\alpha$ is about $0.5$.  For the estimated density of 
the hot core in Orion of $n_H \approx 10^7$ cm$^{-3}$ , this scaling 
implies $<B>$ = $10$ milli-gauss. Zeeman measurement of the magnetic 
fields using OH masers in BN/KL imply magnetic fields as high as 
15 milli-gauss spread across a roughly 30 arcsecond diameter region 
($\sim 10^4$ AU; Cohen et al. 2006).  Similar results are found in 
other massive star forming cores such as Cepheus A (Bartkiewicz et 
al. 2005;  Vlemmings  2006). Thus, we assume that the magnetic 
field had a strength of 10 milli-gauss when a small group of massive 
stars started to form in OMC1 about $10^5$ years ago.

If the star formation efficiency that led to the formation of a 
40 M$_{\odot}$ sub-cluster of massive stars in OMC1 is 40\%, the initial 
mass of gas involved must have been about 100 M$_{\odot}$.  Accretion
of gas from the OMC1 core and dynamical friction enabled the forming 
stars to migrate deep into the cluster potential well to produce a 
compact, non-hierarchical multiple star system. This gravitational 
contraction of gas and stars ultimately provides the energy for the 
launched gas and ejected stars, so it is convenient to express the 
gravitational energy as $E_{48}$, the energy required to power the OMC1 
outflow and run-away stars in units of $10^{48}$ ergs.  The final radius 
is then given by  $r \approx GM^2/E =  178 M_{100}^2 E_{48}^{-1}$ AU, 
where $M_{100}$ is the enclosed mass (in units of 100 M$_{\odot}$).  
We postulate that during the last $10^5$ years,  4 or more stars 
formed a non-hierarchical system with a characteristic diameter of 
order 200 to 400 AU and a total mass of about 40 M$_{\odot}$.  Such 
non-hierarchical multiple star systems are dynamically unstable and 
typically decay within 10 to 100 dynamical times (Sterzik \& 
Durisen 1998).

The orbital motion of the protostars in their mutual gravitational 
potential well and the orbital motion of gas in individual circumstellar 
disks will each contribute to the amplification of the magnetic field 
by means of a shear dynamo.  In linear winding, the final field strength 
$B$ is related to the initial field as $B = 2 \pi N B_0$ , where $N$ is 
the number of windings of the field lines.  When equipartition is 
reached, $B^2 \approx  8 \pi \rho c_s^2$,  where $\rho = \mu m_H n_H$
 = $3 M / 4 \pi r^3$ is the gas density, and $c_s$ is the effective sound 
speed including turbulent motions.  The maximum field strength is 
obtained when the effective sound speed is similar to the gravitational 
escape speed.  Thus, $c_s^2 \approx  GM/r$ where $M$ is the mass enclosed 
inside radius $r$.   The spatially averaged equipartition value of $B$ 
is then given by 
$ B \approx (6 / r^3)^{1/2}  
           = (6 G)^{1/2} M / r^2   \approx   42 E_{48}^{1/2} r_{100}^{-3/2}
           ~   (gauss) 
$
where $r_{100}$ is in units of 100 AU.   The orbital time scale about 
100 M$_{\odot}$ of gas and stars at this radius is about 100 years, 
so to obtain the maximum average field of order 40 gauss from an initial 
field of $B_0 = 10$ milli-gauss with linear growth requires about 700 orbits 
or less than $0.7 \times 10^5$ years, comparable to the typical time required 
to accrete a star.  

Efficient field amplification of $B$ in the potential well requires a 
sufficient degree of ionization to couple the magnetic field to the gas, 
and assumes that field loss due to reconnection or magnetic diffusion is 
unimportant.  Ionization in this scenario can be maintained by a combination 
of  X-ray and energetic particle radiation expected from stars with 
strong magnetic fields, and possibly by the decay of short-lived radioactive 
species thought to be abundant in massive star forming regions
(Diehl et al. 2004, 2006a, b).

Gas tends to be removed from the region occupied by compact multiple star 
systems by accretion onto individual circumstellar disks or expulsion by 
gravitational torques (Lubow \& Artymowicz 1996).  The removal of gas from 
the intermediate region with a characteristic length-scale comparable to 
the average interstellar separation implies that this region will have 
a large Alfven speed.   For  a  field strength of $40$ gauss, the Alfven 
speed exceeds $10^3$  km s$^{-1}$ for a density less than 
$1.3 \times 10^{-14}$ g cm$^{-3}$ ($n_H < 6 \times 10^9$ cm$^{-3}$). 

The momentum in the BN/KL outflow (Snell et al. 1983) is about  
$P_{out} = 160$ M km s$^{-1}$.    The momentum delivered by the 
`magnetic bomb' to the surrounding gas, $P_{MB} = M_{gas} V_A$, must 
therefore be at least this large.  Consequently, the gas mass in the 
magnetized region must satisfy $M_{gas} > 3 P^2_out / B^2 r^3 \approx 0.3$ 
M$_{\odot}$ for  $B$ = 40 gauss and $r$ = 100 AU.  These parameters imply 
an average Alfven speed of about 600  km s$^{-1}$, somewhat greater than 
the velocity of the very fastest knots in the outflow. 

The strongest fields are likely to be generated by shear dynamos in 
the innermost parts of circumstellar disks surrounding individual massive 
stars. Thus, the highest velocity ejecta observed in the BN/KL outflow 
may have been launched by the dynamical disruption of highly magnetized 
material from within a few to tens of AU of the individual stars by 
the penetrating encounter that disrupted the star system. Reconnection of 
magnetic fields, steep magnetic pressure gradients, and centrifugal 
acceleration  may also have contributed to the acceleration of the 
fastest ejecta.  Fast ejecta slamming into slower material farther 
out can produce high velocity bullets and fingers 
(McCaughrean \& Mac Low 1997; Stone, Xu, \& Mundy 1995).   Highly 
perturbed circumstellar disks with radii of at least several AU may 
survive the dynamical interactions and may be dragged along by the 
ejected stars.  These remnant disks may continue to fuel collimated outflow 
activity along axes that reflect the changed disk orientations. 

In summary, we propose that the Orion BN/KL outflow was powered by 
magnetic energy generated by the orbital motion of a non-hierarchical 
multiple stars system of 4 or more stars over $10^4$ to $10^5$ years, 
and released by dynamical decay a mere 500 years ago.  This event led 
either to the formation of one or more compact binaries or possibly 
a stellar merger.  A testable prediction of this scenario is that 
the Galaxy should contain a population of  compact, massive, magnetized 
hot cores with field strengths of order 10 to 100 Gauss.  Future 
interferometric observations of Zeeman splitting of far-infrared, 
sub-millimeter, or millimeter spectral lines should detect these 
magnetic fields on 100 AU length scales.

\acknowledgements 
This work was supported by NSF grant  AST0407356 and the CU Center
for  Astrobiology funded by NASA under Cooperative 
Agreement no. NNA04CC11A issued by the Office of Space Science.

\end{document}